\documentclass[letter]{aa} 

\usepackage{graphicx}
\usepackage{longtable}
\usepackage{txfonts}
\begin{document}

   \title{Signatures of rocky planet engulfment in HAT-P-4\thanks{Based
on observations obtained at the Gemini Observatory, which is
operated by the Association of Universities for Research in Astronomy,
Inc., under a cooperative agreement with the NSF on behalf of the
Gemini partnership: the National Science Foundation (US),
the National Research Council (Canada), CONICYT (Chile), Min. de
Ciencia y Tecnolog\'ia (Argentina), and Ministério da Ciência, Tecnologia e Inovação (Brazil).}
   }
   
\subtitle{Implications for chemical tagging studies}

   
   \titlerunning{Rocky planet engulfment in HAT-P-4}
   \authorrunning{Saffe et al.}

   \author{C. Saffe\inst{1,2,4}, E. Jofr\'e\inst{3,4}, E. Martioli\inst{5}, M. Flores\inst{1,2,4},
   R. Petrucci\inst{3,4} \and M. Jaque Arancibia\inst{1,4}
          }

   \institute{Instituto de Ciencias Astron\'omicas, de la Tierra y del Espacio (ICATE-CONICET), C.C 467, 5400, San Juan, Argentina.
             \email{csaffe,matiasflorestrivigno,mjaque@conicet.gov.ar}
         \and
         Universidad Nacional de San Juan (UNSJ), Facultad de Ciencias Exactas, F\'isicas y Naturales (FCEFN), San Juan, Argentina.
         \and
         Observatorio Astron\'omico de C\'ordoba (OAC), Laprida 854, X5000BGR, C\'ordoba, Argentina.
             \email{emiliano,romina@oac.unc.edu.ar}
         \and
         Consejo Nacional de Investigaciones Cient\'ificas y T\'ecnicas (CONICET), Argentina
         \and
         Laborat\'{o}rio Nacional de Astrof\'{i}sica (LNA/MCTI),  Rua Estados Unidos 154, Itajub\'{a}, MG, Brazil
           \email{emartioli@lna.br}
        }

   \date{Received xxx, xxx ; accepted xxxx, xxxx}

 
  \abstract
  {}
   {To explore the possible chemical signature of planet formation in the binary
   system HAT-P-4, by studying abundance vs condensation temperature T$_{c}$ trends.
   The star HAT-P-4 hosts a planet detected by transits while its stellar companion does not have any detected planet. 
   We also study the Lithium content, which could shed light on the problem of \ion{Li}{} depletion in exoplanet host stars.}
   {We derive, for the first time, both stellar parameters and high-precision chemical abundances by applying a 
   line-by-line full differential approach. The stellar parameters were determined by imposing ionization
   and excitation equilibrium of \ion{Fe}{} lines, with an updated version of the FUNDPAR program, together with ATLAS9
   model atmospheres and the MOOG code. We derived detailed abundances of different species with equivalent widths
   and spectral synthesis with the MOOG program.
   }
   {The exoplanet host star HAT-P-4 is found to be $\sim$0.1 dex more metal rich than its companion,
   which is one of the highest differences in metallicity observed in similar systems.
   This could have important implications for chemical tagging studies, disentangling groups of stars with a common origin.
   We rule out a possible peculiar composition for each star as $\lambda$ Bo\"otis, $\delta$ Scuti or a Blue Straggler. 
   The star HAT-P-4 is enhanced in refractory elements relative to volatile when compared to its stellar companion.
   Notably, the Lithium abundance in HAT-P-4 is greater than in its companion by $\sim$0.3 dex, which is contrary to the
   model that explains the Lithium depletion by the presence of planets.
   We propose a scenario where, at the time of planet formation, the star HAT-P-4 locked the inner refractory
   material in planetesimals and rocky planets, and formed the outer gas giant planet at a greater distance.
   The refractories were then accreted onto the star, possibly due to the migration of the giant planet.
   This explains the higher metallicity, the higher Lithium content, and the negative T$_{c}$ trend detected.
   A similar scenario was recently proposed for the solar twin star HIP 68468, which is in some aspects similar to HAT-P-4.
   We estimate a mass of at least M$_{rock} \sim$ 10 M$_{\oplus}$ locked in refractory material in order to reproduce
   the observed T$_{c}$ trends and metallicity. 
   }
   {}
   

   \maketitle
%

\section{Introduction}

The detection of a possible chemical signature of planet formation in the photospheres
of planet host stars is a challenge for the current studies.
Several authors attempted to detect this signature by looking at the condensation temperature
(T$_{c}$) trend in planet host stars \citep[e.g.][]{gonzalez97,smith01,gratton01}.
In particular, \citet{melendez09} showed that the atmosphere of the Sun is deficient in
refractory\footnote{Refractory and volatile species are those with T$_{c}>$ 900 K and T$_{c}<$ 900 K.}
elements when compared to the average abundances of 11 solar twins, also showing a clear T$_{c}$ trend. They
proposed that the missing refractories are probably locked in terrestrial planets
and rocky material that orbits the Sun. This idea was then supported by \citet{ramirez09} who
studied a sample of 64 solar twins and analogs with and without planets.
On the other hand, other authors suggest that the T$_{c}$ trends depend on the galactic chemical
evolution (GCE), the stellar age or probably the galactic birth place of the stars \citep[e.g.][]{adi14,adi16},
competing with the proposed chemical signature of planet formation \citep[e.g. ][]{gonz-hern13}.

Most multiple and binary stars are believed to have formed coevally from a common molecular cloud.
Wide binaries are particularly valuable, because both components can be presumed to have the
same age and initial composition, greatly diminishing the mentioned age and GCE effects.
Several studies showed that there may be small differences in the chemical composition
of their components, possibly due to the planet formation process \citep[][]{gratton01,desidera04,desidera06}.
The similarity between both components of a binary system made possible to achieve
a higher precision in a differential study \citep[e.g. ][]{saffe15,saffe16}.
To date, more than 2700 planetary systems have been reported\footnote{http://exoplanet.eu/catalog/},
but only a handful of these systems are binaries with similar components.
Three of these remarkable systems were studied in detail: 16 Cyg, HAT-P-1 and HD 80606
\citep{laws-gonzalez01,ramirez11,schuler11,tucci-maia14,liu14,saffe15}.
There are also binary systems with circumstellar planets orbiting both stars of the system \citep[e.g. ][]{mack14}.
These studies show that a T$_{c}$ trend is probably present between the stars of 16 Cyg but not
in the case of HAT-P-1 nor HD 80606. Notably, the binary system $\zeta^{2}$ Ret (where one component
is orbited by a dust disk with no planet detected), also shows a T$_{c}$ trend
between their stars, supporting the chemical signature of planet formation \citep{saffe16}.
Then, there is a clear need of additional systems to be studied through detailed abundance analyses
in order to reach more significant conclusions.

As a result of the planetary transit survey HATNet,
\citet{kovacs07} discovered a giant planet of 0.68 M$_{Jup}$ orbiting the star {HAT-P-4}
(=BD +36 2593) at a distance of 0.04 AU. They estimated a density of 0.41 g cm$^{-3}$,
being one of the lowest density hot Jupiters known.
Then, \citet{mugrauer14} showed that HAT-P-4 forms a wide binary system separated 91.8 arcsec from
its companion (TYC 2569-744-1, hereafter component B), and showed that both stars present very
similar spectra (G0V + G2V).
The estimated separation is 28446 AU, being considered by \citet{mugrauer14} as a 
true binary system based in their common proper motion and similar radial velocity.
To date, there is no planet detected around the B star, being also included in the field G191
(FOV = 8 $\times$ 8 deg) of the HATNet survey \citep{kovacs07}.
As we show in the next sections, the stellar parameters of both stars are very similar,
making this system an ideal case to study the possible chemical signature of planet formation.

Different authors have showed a possible excess of \ion{Li}{} depletion in stars with planets, when compared to stars
without planets \citep{king97,israelian04,israelian09,delgado15,gonzalez14,gonzalez15}.
The \ion{Li}{} depletion was first attributed to the presence of planets, by possibly increasing the angular momentum
of the star (e.g. during planet migration) and increasing its convective mixing \citep{israelian09}.
On the other hand, some works show that the \ion{Li}{} depletion is related to a bias in age,
mass and metallicity, and not due to the presence of planets nor the planet formation process
\citep{luck-heiter06,baumann10,ramirez12,carlos16}.
The study of similar stars in binary systems can help to understand the origin of \ion{Li}{} depletion.
In the 16 Cyg binary system, the B component hosts a planet of 1.5 M$_{Jup}$ \citep{cochran97}
and presents a \ion{Li}{} content $\sim$4 times lower than the A star \citep{king97},
supporting the model of \ion{Li}{} depletion by the presence of planets. However, \citet{ramirez11} showed
that the slightly different mass of stars A and B could explain its different \ion{Li}{} content i.e. 16 Cyg
in principle can support both scenarios. 
Notably, HAT-P-4 is in some way complementary to 16 Cyg: in this case, the planet host star presents
the higher \ion{Li}{} content, which is contrary to the model of \ion{Li}{} depletion due to the presence of planets.
Then, it is worthwhile exploring both the T$_{c}$ trends and the \ion{Li}{} depletion in this remarkable system.

\section{Observations and data reduction}

Observations of HAT-P-4 binary system were acquired through GRACES
(Gemini Remote Access to CFHT ESPaDOnS Spectrograph).
This device takes advantage of the high-resolution 
ESPaDOnS\footnote{Echelle SpectroPolarimetric Device for the Observation of Stars}
spectrograph, located at the Canada-France-Hawaii Telescope (CFHT) and fed by an
optical fiber connected to the 8.1 m Gemini North telescope at Mauna Kea, Hawaii.
We used the 1-fiber object-only observing mode which provides an average
resolving power of $\sim$67500 between 4500 and 8500 \AA\footnote{http://www.gemini.edu/sciops/instruments/graces/spectroscopy/spectral-range-and-resolution}.
The stellar spectra were obtained under a Fast Turnaround (FT) observing
mode requested to the Gemini Observatory (program ID: GN-2016A-FT-25, PI:
C. Saffe). The observations were taken on June 3, 2016, with the B star observed
immediately after the A star, using for both the same spectrograph configuration.
The exposure times were 2 $\times$ 17 min on each target, obtaining a final signal-to-noise
ratio of S/N $\sim$400 measured at $\sim$6000 {\AA} in the combined spectra.
The final spectral coverage is 4050-10000 \AA.
The Moon was also observed with the same spectrograph set-up, achieving a similar S/N
to acquire the solar spectrum as initial reference.
GRACES spectra were reduced using the code OPERA\footnote{Open source Pipeline for ESPaDOnS Reduction
and Analysis} \citep{martioli12}. More recent documentation on OPERA can be found
at the ESPECTRO project webpage\footnote{http://wiki.lna.br/wiki/espectro}.

\section{Stellar parameters and abundance analysis}

\begin{figure}
\centering
\includegraphics[width=6cm]{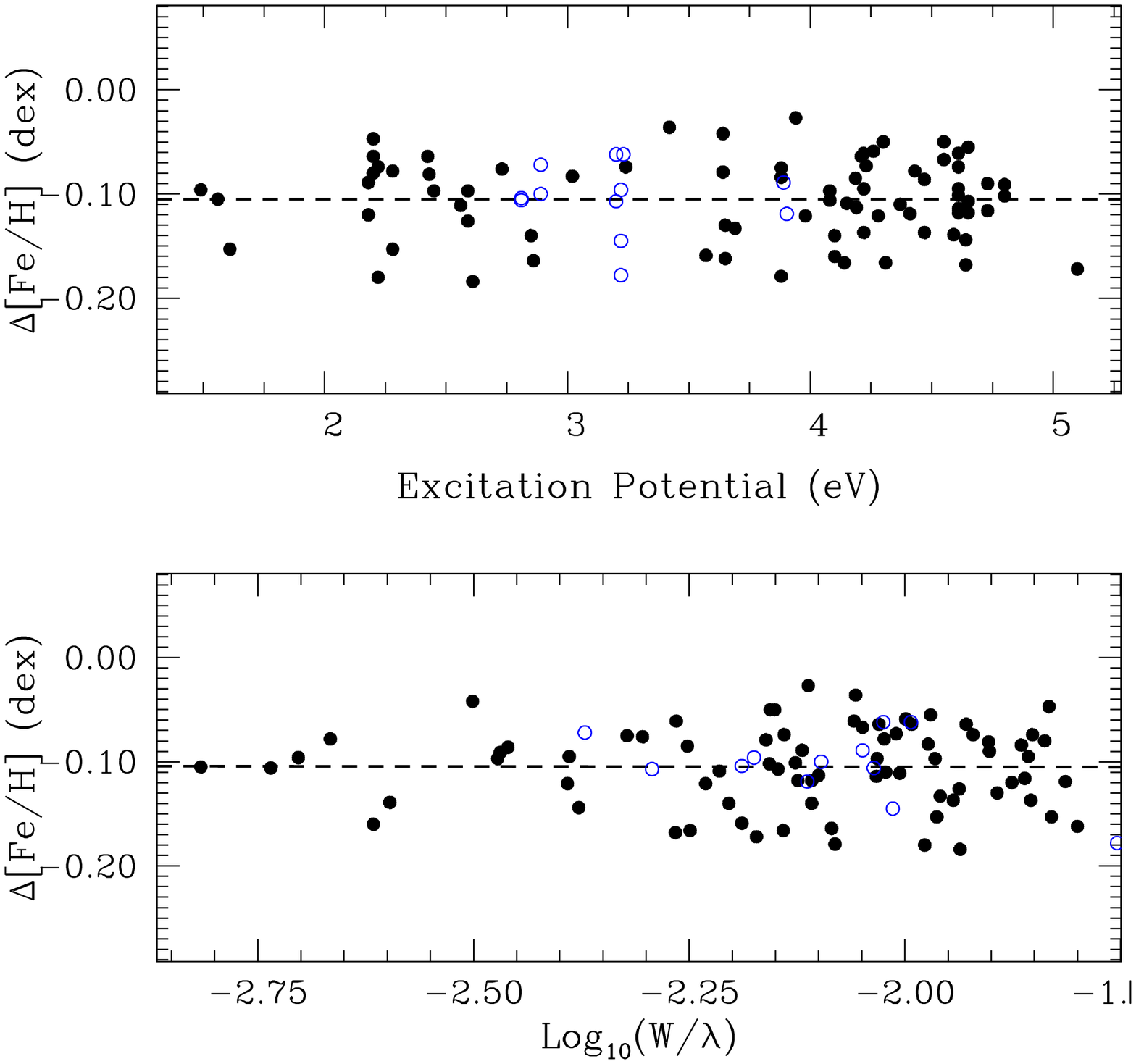}
\caption{Differential abundance vs excitation potential (upper panel) 
and vs reduced EW (lower panel) for the case (B - A).
Filled and hollow circles correspond to \ion{Fe}{I} and \ion{Fe}{II}.}
\label{equil-relat}%
\end{figure}

Fundamental parameters (T$_{eff}$, {log $g$}, [Fe/H], v$_{turb}$)
were derived following a similar procedure to our previous work \citep{saffe15,saffe16}.
We measured the equivalent widths (EW) of \ion{Fe}{I} and \ion{Fe}{II} lines by
using the IRAF task splot, and then continued with other chemical species.
Both the lines list and relevant laboratory data were taken from literature \citep{liu14,melendez14,bedell14}. 
We imposed excitation and ionization balance of \ion{Fe}{} lines, using the differential
version of the FUNDPAR program \citep{saffe11,saffe15}. This code made use of ATLAS9 model atmospheres
\citep{kurucz93} together with the 2014 version of the MOOG program \citep{sneden73}.
Stellar parameters were determined using the Sun as reference i.e. (A - Sun) and (B - Sun),
by adopting (5777 K, 4.44 dex, 0.00 dex, 0.91 km/s) for the Sun\footnote{We estimated
a v$_{turb}$ of 0.91 km/s for the Sun by requiring zero slope between \ion{Fe}{I} abundances and EW$_{r}$}.
Then, we recalculated the parameters using the A star as reference i.e. (B - A), by adopting
for the A star the parameters derived for the case (A - Sun) (see Table \ref{stellar.params}).
The errors in the stellar parameters were derived following the procedure detailed in \citet{saffe15},
which takes into account the individual and the mutual covariance terms of the error propagation.
Within the errors, the parameters of the B star agree using the Sun or the A star as reference.
We note that the A star is $\sim$0.1 dex more metal rich than the B star.
Figure \ref{equil-relat} shows abundance vs excitation potential (top panel)
and abundance vs EW$_{r}$ (bottom panel), both for the case (B - A).

\begin{table}
\centering
\caption{Stellar parameters derived for each star.}
\vskip -0.1in
\scriptsize
\begin{tabular}{ccccc}
\hline
\hline
 (Star - Reference) & T$_{eff}$ & log $g$ & [Fe/H] & v$_{turb}$ \\
  & [K] & [dex] & [dex] & [km s$^{-1}$] \\
\hline
(A - Sun)              &  6036 $\pm$ 46 & 4.33 $\pm$ 0.13 &  0.277 $\pm$ 0.007 & 1.29 $\pm$ 0.07 \\
(B - Sun)              &  6037 $\pm$ 37 & 4.38 $\pm$ 0.14 &  0.175 $\pm$ 0.006 & 1.21 $\pm$ 0.07 \\
(B - A)                &  6035 $\pm$ 36 & 4.39 $\pm$ 0.10 & -0.105 $\pm$ 0.006 & 1.22 $\pm$ 0.06 \\
\hline
\end{tabular}
\normalsize
\label{stellar.params}
\end{table}

The next step was the derivation of abundances for all remaining chemical elements,
The hyperfine structure splitting (HFS) was considered for 
\ion{V}{I}, \ion{Mn}{I}, \ion{Co}{I}, \ion{Cu}{I} and \ion{Ba}{II} using the HFS
constants of \citet{kurucz-bell95} and performing spectral synthesis for these species.
We also derived the \ion{Li}{I} abundance by using spectral synthesis with the resonance line 6707.80 {\AA}
which includes the doublet 6707.76 {\AA}, 6707.91 {\AA} and HFS components.
We corrected \ion{Na}{I} abundance by NLTE (Non-Local Thermodynamic Equilibrium) effects, interpolating
in the data of \citet{shi04} and adopting Na(NLTE) - Na(LTE) $\sim$ -0.07 dex for each star.
We also applied NLTE corrections to \ion{O}{I} (-0.18 dex and -0.17 dex for the A and B stars),
by interpolating in the data of \citet{ramirez07}.
These corrections are relative to the Sun, which implies that NLTE effects for the case (B - A)
are not significative given the high similarity between the stars A and B.
\vskip -0.5in

\section{Results and discussion}

\begin{figure}
\centering
\includegraphics[width=7cm]{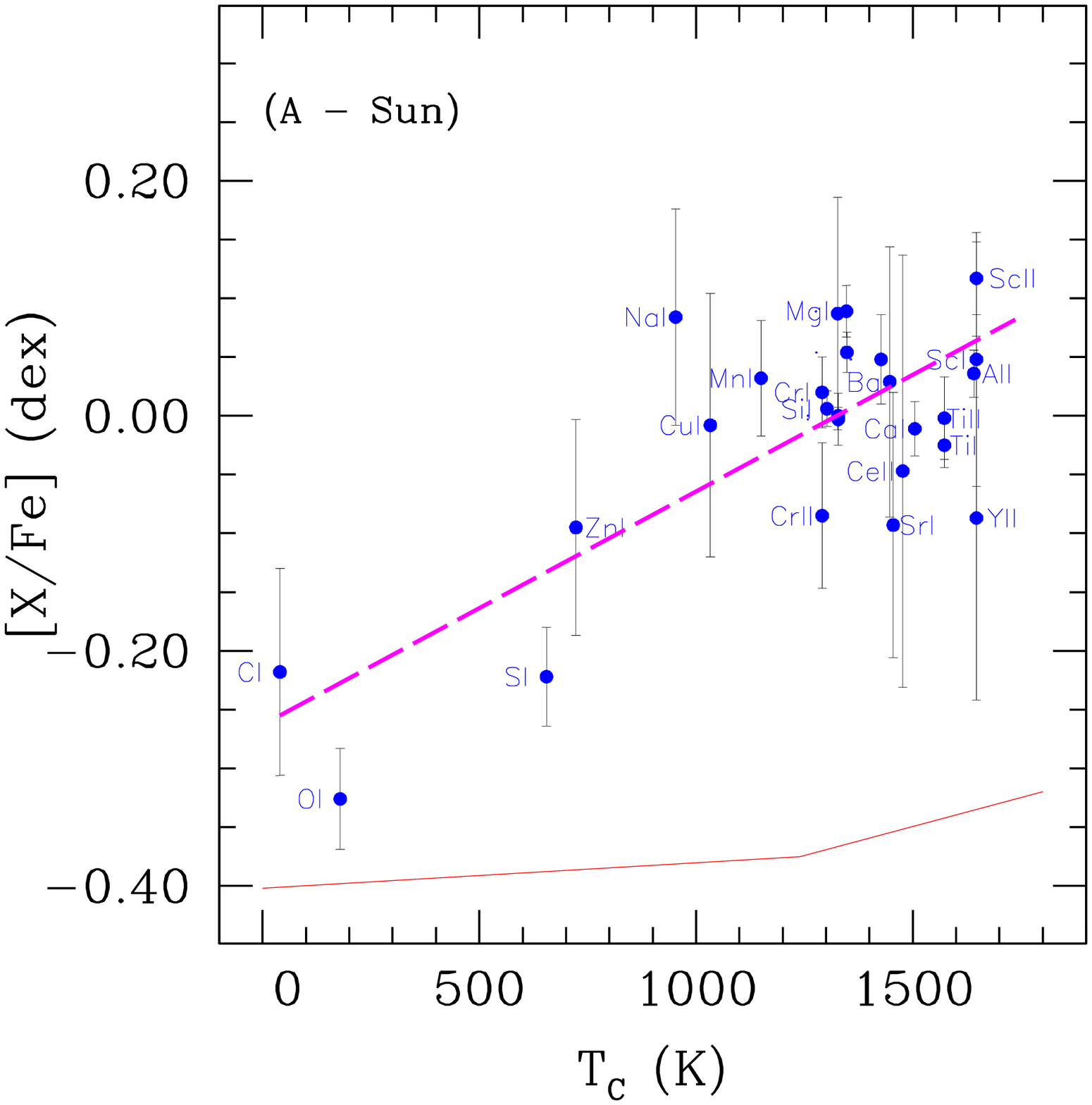}
\caption{Differential abundances vs T$_{c}$ for (A - Sun).
The long-dashed line is a weighted linear fit to the differential abundance values,
while the continuous line shows the solar-twins trend of \citet{melendez09}.}
\label{abund-HAT-sun}%
\end{figure}

Condensation temperatures were taken from the 50\% T$_{c}$ values derived by \citet{lodders03}
for a solar system gas with [Fe/H]=0. As suggested by the referee, it would be helpful the calculation
of other T$_{c}$ sequences for different metallicity values.
We corrected by GCE effects for the case (star - Sun) but not for the case (B - A),
by adopting the GCE fitting trends of \citet{gonz-hern13} i.e. following the same procedure of \citet{saffe15}.
Figure \ref{abund-HAT-sun} presents the corrected abundance values vs T$_{c}$ for
(A - Sun)\footnote{For the B star, the values are similar to those of Figure \ref{abund-HAT-sun}.}.
Table \ref{slopes} presents the slopes and uncertainties of the linear fits.
The positive slopes for the case (star - Sun) indicate a higher content of refractories
(those with T$_{c}>$ 900 K) relative to volatiles (T$_{c}<$ 900 K).

\begin{table}
\centering
\caption{Derived slopes (abundance vs T$_{c}$) and their uncertainties.
We also included the V magnitude and mass of the stars.
}
\vskip -0.1in
\scriptsize
\begin{tabular}{rrcc}
\hline
\hline
(Star - Reference)  & Slope$\pm \sigma$  & V & Mass \\
  & [10$^{-5}$ dex/K] & [mag] & [M$_{\sun}$] \\
\hline
(A - Sun)                      & +19.88 $\pm$ 2.29 & V$_{A}=$ 11.12 & M$_{A}=$ 1.24 $\pm$ 0.06  \\
(B - Sun)                      & +14.59 $\pm$ 1.93 & V$_{B}=$ 11.38 & M$_{B}=$ 1.17 $\pm$ 0.05  \\
(B - A)                        &  -5.18 $\pm$ 1.15  \\
(B - A)$_{Refr}$               &  -7.81 $\pm$ 2.61  \\
\hline
\end{tabular}
\normalsize
\label{slopes}
\end{table}

\begin{figure}
\centering
\vskip -0.2in
\includegraphics[width=7cm]{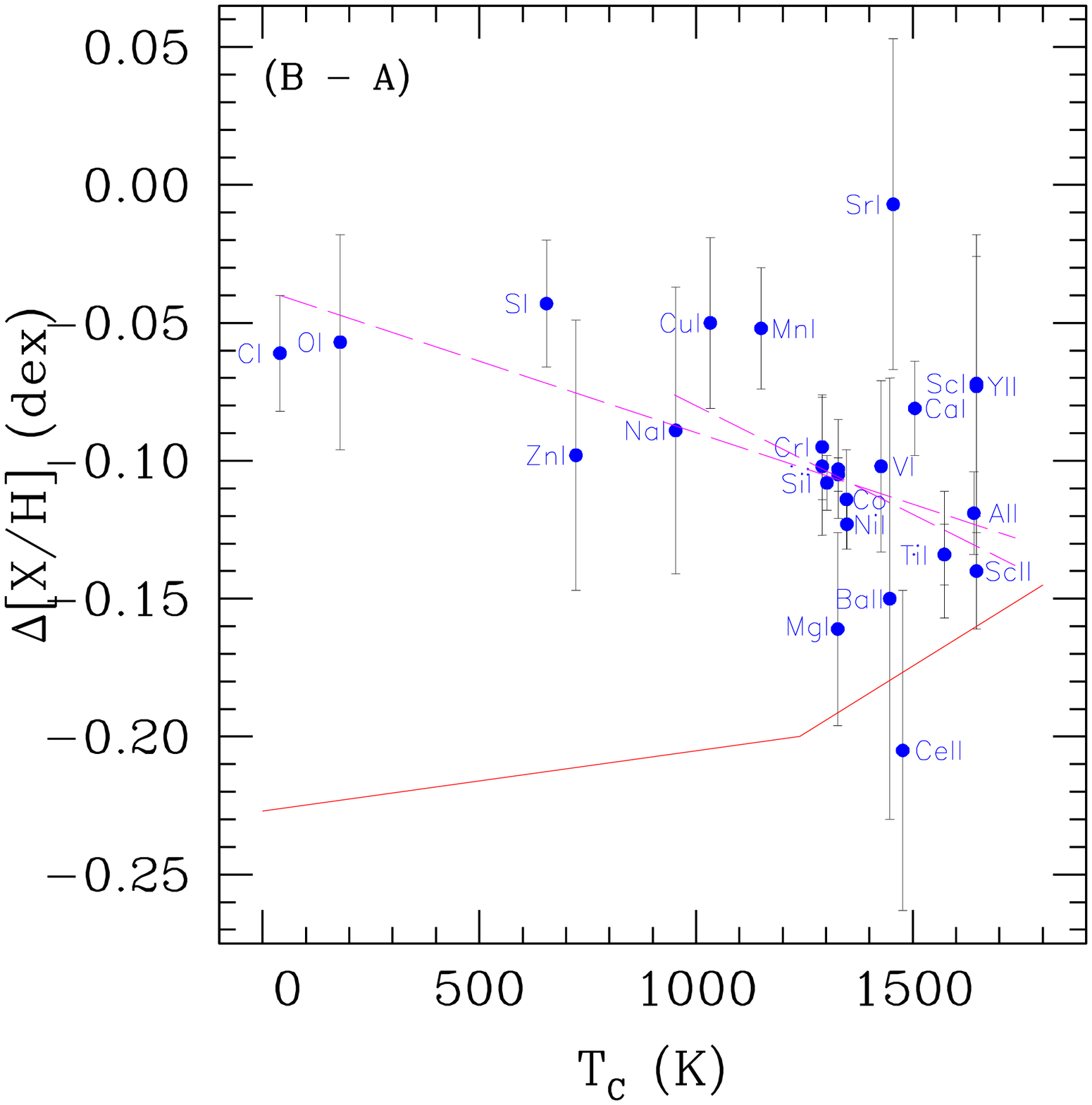}
\caption{Differential abundances (B - A) vs T$_{c}$.
Long-dashed lines are weighted linear fits to all species and to refractory species.
The solar-twins trend of \citet{melendez09} is shown with a continuous line.}
\label{relat.tc}%
\end{figure}

The differential abundances for (B - A) are presented in Figure \ref{relat.tc}.
The continuous line in this Figure presents the solar-twins trend of \citet{melendez09} (vertically shifted).
Long-dashed lines are weighted linear fits to all elements and to refractory elements,
showing both similar negative slopes (see also Table \ref{slopes}).
Some species such as \ion{Sr}{I} and \ion{Ce}{II} seem to possibly drive the trends, however
we obtained very similar slopes after excluding these species.
The average refractory and volatile [X/H] abundance values for (B - A) are {-0.105 $\pm$ 0.007 dex} and
{-0.065 $\pm$ 0.015 dex}, which together with the negative slopes of Figure \ref{relat.tc}
points toward a higher content of refractories in the A star than in its stellar companion.

\subsection{Do HAT-P-4 stars have a peculiar composition?}

We detected a difference of $\sim$0.1 dex in metallicity between the stars A and B,
which is one of the highest differences found in similar systems
\citep[e.g. ][]{desidera04,desidera06}. In this section we discuss the possible chemical pattern of both stars.
$\lambda$ Bo\"otis stars show moderate surface underabundances of most Fe-peak elements, but solar
abundances of C, N, O and S \citep[e.g.][]{paunzen04}, in contrast to the metal-rich content of stars A and B.
   
$\delta$ Scuti stars are pulsating variables with $\sim$A6-F6 spectral types
and mass between 1.4 - 3.0 M$_{\sun}$ \citep[see e.g. the catalogs of ][]{rodriguez00,liakos-niarchos17}.
We determined the stellar masses through a Bayesian method using the
PARSEC\footnote{http://stev.oapd.inaf.it/cgi-bin/param\_1.3} isochrones \citep{bressan12},
obtaining 1.24 $\pm$ 0.06 M$_{\sun}$ and 1.17 $\pm$ 0.05 M$_{\sun}$ for the stars A and B.
Both the temperature and mass of A and B are lower than all $\delta$ Scuti stars of these catalogs.
One of the coolest $\delta$ Scuti stars with abundance determination is CP Boo \citep[6320 K, ][]{galeev12},
which is $\sim$300 K hotter than the A star. However, the A star is enhanced ($\gtrsim$ 0.20 dex) in 
\ion{Na}{I}, \ion{Mg}{I}, \ion{Al}{I}, \ion{Sc}{II}, \ion{V}{I},
and strongly enhanced ($>$ 0.40 dex) in \ion{Sr}{II} and \ion{Y}{II} compared to CP Boo.
Also, the stars A and B lie out of the instability strip boundaries 
\citep[e.g. Figures 6 and 7 of ][]{liakos-niarchos17} and no stellar pulsations have been reported.

Blue stragglers (BS) are stars significantly bluer than the main-sequence turnoff of the 
cluster (or population) to which they belong.
There are binaries where one component is BS, being impossible to fit both
components with a single isochrone \citep{desidera07}. However, in our system the ages of both stars
agree within the errors (2.7 $\pm$ 1.3 Gyr and 2.9 $\pm$ 1.8 Gyr),
also derived with the PARSEC isochrones \citep{bressan12}.
BS stars present significant rotational velocities, intense activity and very low \ion{Li}{} content
\citep[e.g. ][]{fuhrmann-bernkopf99,schirbel15,ryan01}. None of these characteristics
are seen in the stars A or B.

There are no firm reasons to identify any component of the binary system as peculiar.
Then, a possible different accretion would be the most plausible explanation
for the chemical differences observed.
We estimated the rocky mass of depleted material in the A star following \citet{chambers10}.
Adopting a convection zone similar to the present Sun (M$_{cz}$ = 0.023 M$_{\sun}$) we obtain
M$_{rock}\sim$10 M$_{\oplus}$. However, adopting a higher M$_{cz}$ value (e.g. 0.050 M$_{\sun}$)
at the time of planet formation, we derive M$_{rock}\sim$20 M$_{\oplus}$.
\citet{cody-sasselov05} found that for stars with 1.1 M$_{\sun}$, a polluted convection zone is slightly
smaller ($\sim$1 \%) than its unpolluted counterpart, however the trend is reversed for stars with M $<=$ M$_{\sun}$.
More recently, \citet{vansaders12} found that the acoustic depth to the base of the CZ varies at the 0.5–1 \% per 0.1 dex
level in [Z/X]. A change of 1\% in M$_{cz}$ translates into $\sim$0.5 M$_{\oplus}$ of derived accreted material.
Then, the estimation of at least 10 M$_{\oplus}$ of accreted material should be considered as a first order
approximation.

\subsection{The hot Jupiter planet and the Lithium content}

The formation of hot Jupiter planets is mainly explained by the model of core accretion \citep{pollack96},
and by gravitational instability \citep{boss00}.
In both cases, some kind of migration from the original location is required to explain the current
orbit at $\sim$0.04 AU, with the possible accretion of inner planets into the star \citep[e.g. ][]{mustill15}.
Other posibility is the "in situ" formation of the hot Jupiter \citep[e.g. ][]{bodenheimer00}, without migration
from greater distances. However, abundances of C/O and O/H ratios suggest that some hot Jupiters originate
beyond the snow line \citep{brewer17}. With the present data, it is difficult to disentangle between the
different formation scenarios. Additional planets were searched in the A star by transits \citep{smith09,ballard11}
and radial velocity \citep{knutson14}, with no success. Then, a scenario with a possible migration and accretion
could not be discarded.

\begin{figure}
\centering
\vskip -0.2in
\includegraphics[width=5cm]{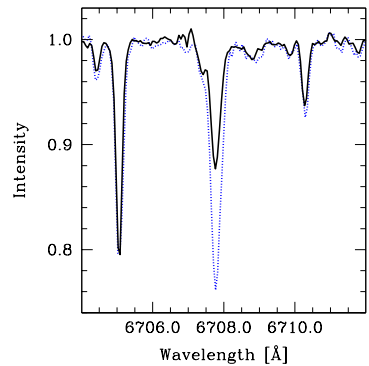}
\caption{Stellar spectra near the Lithium line 6707.8 \AA\ for the A (blue dotted line) and
B (black continuous line) stars.}
\label{lithium}%
\end{figure}

Figure \ref{lithium} shows that the line \ion{Li}{I} 6707.8 \AA\ is stronger in the A star
(blue dotted line) compared to B (black continuous line),
with abundances of 1.47 $\pm$ 0.05 and 1.17 $\pm$ 0.04 dex.
In this binary system, the star which hosts the planet also presents the higher
\ion{Li}{} content, in contrast to the model that explains the \ion{Li}{} depletion by the presence of planets.
A similar rotational velocity (7.0 $\pm$ 0.9 and 6.1 $\pm$ 1.0 km s$^{-1}$ estimated from our spectra),
suggests that the rotational mixing is not the cause for the different \ion{Li}{} content.
The sligthly different masses of stars A and B suggest a difference of 0.2-0.3 dex
in the \ion{Li}{} abundances, by fitting general literature trends
(e.g. Figure 3 of \citet{delgado15} and Figure 9 of \citet{ramirez12}).
This can explain (at least in part) the higher \ion{Li}{} content, in agreement with the scenario proposed by
\citet{ramirez11} for 16 Cyg. 
Another possible scenario is the accretion of material in the A star, which could increase 
its \ion{Li}{} abundance and agree with its higher metal content.

\subsection{On the relative T$_{eff}$ difference}

\begin{figure}
\centering
\includegraphics[width=6cm]{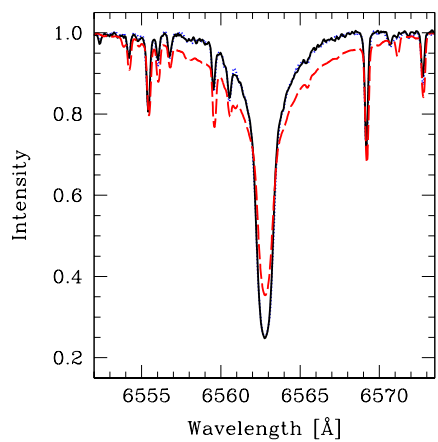}
\caption{Spectral region near the H$\alpha$ line for the stars A and B (blue dotted and 
black continuous line). A synthetic spectra (red dashed line) with T$_{eff}$(A) - 150 K = 5886 K
is also showed for comparison.}
\label{Halpha}
\end{figure}

In this section we discuss about the relative T$_{eff}$ difference between the stars A and B.
Balmer lines could be used as a T$_{eff}$ proxies for solar-type stars \citep[e.g. ][]{barklem02}.
Figure \ref{Halpha} presents the region near the H$\alpha$ line for the stars A (blue dotted line)
and B (black continuous line) almost superimposed, together with a synthetic spectra for comparison
(red dashed line) calculated with T$_{eff}$ = T$_{eff}$(A) - 150 K = 5886 K.
This suggest that stars A and B present a very similar T$_{eff}$.
We also used the photometric calibration of \citet{casa10} with the colors (B-V)$_{T}$ and (J-Ks)
from the Tycho and 2MASS catalogs, estimating a difference of +95 K and -5 K for (B - A).
However, we note that the errors reported for V$_{T}$ and B$_{T}$ vary between 0.07-0.13 mag,
while the errors in JHK are notably lower, 0.01-0.03 mag.

\begin{figure}
\centering
\vskip -0.2in
\includegraphics[width=6cm]{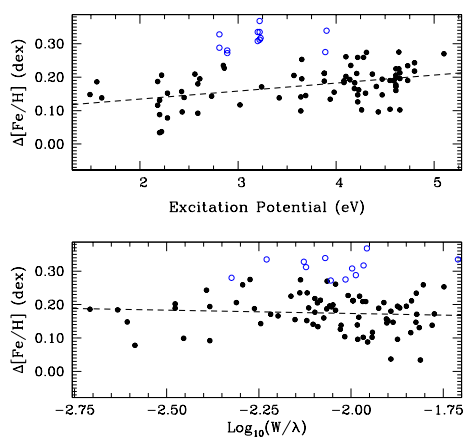}
\caption{Differential abundance vs excitation potential (upper panel) 
and vs reduced EW (lower panel) for the A star, adopting T$_{eff}$(A) - 150 K = 5886 K.
Filled and hollow circles correspond to \ion{Fe}{I} and \ion{Fe}{II}.}
\label{teff.plus}%
\end{figure}

We also explored how the relative T$_{eff}$ difference could change the abundance difference.
For the case (B - A) we varied T$_{eff}$ by -50, -100 and -150 K for the reference A star
and recomputed new solutions for (B - A). In this way, we derived relative [\ion{Fe}{}/\ion{H}{}]
differences between B and A of +0.11, +0.04 and -0.01 dex.
This shows that a T$_{eff}$ difference near $\sim$150 K would be required to remove the abundance
offset between the stars A and B. However, Figure \ref{Halpha} shows that a synthetic spectra
calculated for T$_{eff}$ = T$_{eff}$(A) - 150 K = 5886 K does not fit the observed H$\alpha$ profiles
and does not respect excitation and ionization equilibrium of \ion{Fe}{} lines (Figure \ref{teff.plus}).
Then, we do not find a clear reason to assume such a difference in T$_{eff}$ between the stars A and B.

\subsection{Possible interpretations of the data}

We propose that at the time of planet formation, the A star locked
the orbiting refractory material in planetesimals and rocky planets, and formed the
gas giant in the external disk. This is followed by the accretion of most of the refractories
(possibly due to migration of the gas giant) onto the A star.
This scenario explains the higher metallicity of the A star (no chemically peculiar patterns
are seen),
the higher refractory abundances and the higher \ion{Li}{} content. It is also supported
by the slightly higher mass of the A star (corresponding to a lower convective mass
and lower mixing of the accreted material), and by the non-detection of additional inner planets.
A very similar scenario was recently proposed for the solar twin star HIP 68468 \citep{melendez17},
where the authors suggest that the possible inward migration of a close-in giant planet 
likely produces the engulfment of the inner planetary material by the star. This would explain
the high \ion{Li}{} content (0.6 dex greater than expected for its age), a refractory
enhancement compared to the Sun and a T$_{c}$ trend.
We also note that chemical tagging studies ussually claim to be able to disentangle groups
of stars with common origin within $\sim$0.04 dex of precision \citep[e.g.][]{hogg16}.
Then, an intrinsic difference of $\sim$0.1 dex in metallicity would imply a great challenge
for these works.

\begin{acknowledgements}
The authors thank Drs. R. Kurucz and C. Sneden for making their codes available to us.
E.J., R.P., M.F., and M.J.A. acknowledge the financial support from CONICET in the forms of Post-Doctoral Fellowships.
We also thank the referee Dr. Ivan Ram\'irez for constructive comments that greatly improved the paper.
\end{acknowledgements}


\vskip -0.3in
\begin{thebibliography}{}
\bibitem[Adibekyan et al. (2014)]{adi14} Adibekyan, V., Gonz\'alez Hern\'andez. J., Delgado-Mena, E., et al., 2014, A\&A 564, 15
\bibitem[Adibekyan et al. (2016)]{adi16} Adibekyan, V., Delgado-Mena, E., Figueira, P., et al., 2016, A\&A 592, A87
\bibitem[Ballard et al. (2011)]{ballard11} Ballard, S., Christiansen, J., Charbonneau, et al., 2011, ApJ 732, 41
\bibitem[Barklem et al. (2002)]{barklem02} Barklem, P. S., Stempels, H. C., Allende Prieto, C. et al., 2002, A\&A 385, 951
\bibitem[Baumann et al. (2010)]{baumann10} Baumann, P., Ram\'irez, I., Mel\'endez, J., et al., 2010, A\&A 519, A87
\bibitem[Bedell et al. (2014)]{bedell14} Bedell, M., Mel\'endez, J., Bean, J., et al., 2014, AJ 795, 23
\bibitem[Boss (2000)]{boss00} Boss, A., 2000, ApJ 536, 101
\bibitem[Bodenheimer et al. (2000)]{bodenheimer00} Bodenheimer, P., Hubickyj, O., Lissauer, J., 2000, Icarus 143, 2
\bibitem[Bressan et al. (2012)]{bressan12} Bressan, A., Marigo, P., Girardi, L., et al., 2012, MNRAS 427, 127
\bibitem[Brewer et al. (2017)]{brewer17} Brewer, J., Fischer, D., Madhusudhan, N., 2017, AJ 153, 83
\bibitem[Carlos et al. (2016)]{carlos16} Carlos, M., Nissen, P., Mel\'endez, M., 2016, A\&A 587, A100
\bibitem[Casagrande et al. (2010)]{casa10} Casagrande, L., Ram\'irez, I., Mel\'endez, J., et al., 2010, A\&A 512, A54
\bibitem[Chambers (2010)]{chambers10} Chambers, J., 2010, AJ 724, 92
\bibitem[Cochran et al. (1997)]{cochran97} Cochran, W., Hatzes, A., Butler, P., Marcy, G., 1997, ApJ 483, 457
\bibitem[Cody \& Sasselov (2005)]{cody-sasselov05} Cody, A. M., Sasselov, D., 2005, AJ 622, 704
\bibitem[Delgado-Mena et al. (2015)]{delgado15} Delgado-Mena, E., Bertan de Lis, S., Adibekyan, V., et al., 2015, A\&A 576, 69
\bibitem[Desidera et al. (2004)]{desidera04} Desidera, S., Gratton, R. G., Scuderi, S., et al., 2004, A\&A 420, 683
\bibitem[Desidera et al. (2006)]{desidera06} Desidera, S., Gratton, R. G., Lucatello, S., et al., 2006, A\&A 454, 581
\bibitem[Desidera et al. (2007)]{desidera07} Desidera, S., Gratton, R. G., Lucatello, S., et al., 2007, A\&A 462, 1039
\bibitem[Fuhrmann \& Bernkopf (1999)]{fuhrmann-bernkopf99} Fuhrmann, K., Bernkopf, J., 1999, A\&A 347, 897
\bibitem[Galeev et al. (2012)]{galeev12} Galeev, A., Ivanova, D., Shimansky, V., et al., 2012, Astron. Rep. 56, 11, pp 850
\bibitem[Gonzalez (1997)]{gonzalez97} Gonzalez, G., 1997, MNRAS 285, 403
\bibitem[Gonzalez (2014)]{gonzalez14} Gonzalez, G., 2014, MNRAS 441, 1201
\bibitem[Gonzalez (2015)]{gonzalez15} Gonzalez, G., 2015, MNRAS 446, 1020
\bibitem[Gonz\'alez Hern\'andez et al. (2013)]{gonz-hern13} Gonz\'alez Hern\'andez, J., Delgado-Mena, E., et al., 2013, A\&A 552, A6
\bibitem[Gratton et al. (2001)]{gratton01} Gratton, R. G., Bonanno, G., Claudi, et al., 2001, A\&A 377, 123
\bibitem[Hogg et al. (2016)]{hogg16} Hogg, D., Casey, A., Ness, M., et al., 2016, ApJ 833, 262
\bibitem[Israelian et al. (2004)]{israelian04} Israelian, G., Santos, N., Mayor, M., et al., 2004, A\&A 414, 601
\bibitem[Israelian et al. (2009)]{israelian09} Israelian, G., Delgado-Mena, E., Santos, N., et al., 2009, Nature 462, 12
\bibitem[King et al. (1997)]{king97} King, J., Deliyannis, C., Hiltgen, D., et al., 1997, AJ 113, 1871
\bibitem[Kov\'acs et al. (2007)]{kovacs07} Kov\'acs, G., Bakos, G., Torres, et al., 2007, ApJ 670, L41
\bibitem[Knutson et al. (2014)]{knutson14} Knutson, H., Fulton, B., Montet, B., et al., 2014, ApJ 785, 126
\bibitem[Kurucz (1993)]{kurucz93} Kurucz, R. L. 1993, ATLAS9 Stellar Atmosphere Programs and 2 km/s grid, Kurucz CD-ROM 13, Smithsonian Astrophysical Obs., Cambridge, MA.
\bibitem[Kurucz \& Bell (1995)]{kurucz-bell95} Kurucz, R., Bell, B., 1995, Atomic Line Data, Kurucz CD-ROM 23, Smithsonian Astrophysical Obs., Cambridge, MA.
\bibitem[Laws \& Gonzalez (2001)]{laws-gonzalez01} Laws, C., Gonzalez, G., 2001, ApJ 553, 405
\bibitem[Liakos \& Niarchos (2017)]{liakos-niarchos17} Liakos, A., Niarchos, P., 2017, MNRAS 465, 1181
\bibitem[Liu et al. (2014)]{liu14} Liu, F., Asplund, M., Ram\'irez, I., et al., 2014, MNRASL 442, L51
\bibitem[Lodders (2003)]{lodders03} Lodders, K., 2003, AJ 591, 1220
\bibitem[Luck \& Heiter (2006)]{luck-heiter06} Luck, R., Heiter, U., 2006, AJ 131, 3069
\bibitem[Mack et al. (2014)]{mack14} Mack, C., Schuler, S., Stassun, K., et al., 2014, ApJ 787, 98
\bibitem[Maldonado et al. (2015)]{maldonado15} Maldonado, J., Eiroa, C., Villaver, E., et al., 2015, A\&A 579, A20
\bibitem[Martioli et al. (2012)]{martioli12} Martioli, E., Teeple, D., Manset, N., et al., 2012, Software and Cyberinfrastructure for Astronomy II. Proc. of the SPIE, Vol. 8451, 21 pp
\bibitem[Mel\'endez et al. (2009)]{melendez09} Mel\'endez, J., Asplund, M., Gustafsson, B., et al., 2009, AJ 704, L66
\bibitem[Mel\'endez et al. (2014)]{melendez14} Mel\'endez, J., Ram\'irez, I., Karakas, A., et al., 2014, AJ 791, 14 
\bibitem[Mel\'endez et al. (2017)]{melendez17} Mel\'endez, J., Bedell, M., Bean, J., et al., 2017, A\&A 597, 34
\bibitem[Mugrauer et al. (2014)]{mugrauer14} Mugrauer, M., Ginski, C., Seeliger, M., 2014, MNRAS 439, 1063
\bibitem[Mustill et al. (2015)]{mustill15} Mustill, A., Davies, M., Johansen, A., 2015, ApJ 808, 14
\bibitem[Paunzen (2004)]{paunzen04} Pauzen, E., 2004, The A-Star Puzzle, J. Zverko, J. Ziznovsky, S. Adelman, and W. Weiss Eds., IAU Symp. 224, Cambridge Univ. Press, 443-450
\bibitem[Pollack et al. (1996)]{pollack96} Pollack, J., Hubickyj, O., Bodenheimer, P., et al., 1996, Icarus 124, 62
\bibitem[Ram\'irez et al. (2007)]{ramirez07} Ram\'irez, I., Allende Prieto, C., Lambert, D., 2007, A\&A 465, 271
\bibitem[Ram\'irez et al. (2009)]{ramirez09} Ram\'irez, I., Mel\'endez, J., Asplund, M., 2009, A\&A 508, L17
\bibitem[Ram\'irez et al. (2011)]{ramirez11} Ram\'irez, I., Mel\'endez, J., Cornejo, D., et al., 2011, ApJ 740, 76
\bibitem[Ram\'irez et al. (2012)]{ramirez12} Ram\'irez, I., Fish, J., Lambert, D., et al., 2012, ApJ 756, 46
\bibitem[Rodr\'iguez et al. (2000)]{rodriguez00} Rodr\'iguez, E., L\'opez-Gonz\'alez, M., L\'opez de Coca, P., 2000, A\&AS 144, 469
\bibitem[Ryan et al. (2001)]{ryan01} Ryan, S., Beers, T., Kajino, T., Rosolankova, K., 2001, AJ 547, 231
\bibitem[Saffe (2011)]{saffe11} Saffe, C., 2011, RMxAA 47, 3 
\bibitem[Saffe et al. (2015)]{saffe15} Saffe, C., Flores, M., Buccino, A., 2015, A\&A 582, A17
\bibitem[Saffe et al. (2016)]{saffe16} Saffe, C., Flores, M., Jaque Arancibia, et al., 2016, A\&A 588, A81
\bibitem[Schirbel et al. (2015)]{schirbel15} Schirbel, L., Mel\'endez, J., Karakas, A., et al., 2015, A\&A 584, A116
\bibitem[Shi et al. (2004)]{shi04} Shi, J. R., Gehren, T., Zhao, G., 2004, A\&A 423, 683
\bibitem[Schuler et al. (2011)]{schuler11} Schuler, S., Cunha, K., Smith, V., et al., 2011, ApJL 737, L32
\bibitem[Smith et al. (2001)]{smith01} Smith, V., Cunha, K., Lazzaro, D., 2001, AJ 121, 3207
\bibitem[Smith et al. (2009)]{smith09} Smith, A. M., Hebb, L., Collier Cameron A., et al., 2009, MNRAS 398, 1827
\bibitem[Sneden (1973)]{sneden73} Sneden, C., 1973, ApJ 184, 839 
\bibitem[Tucci Maia et al. (2014)]{tucci-maia14} Tucci Maia, M., Mel\'endez, J., Ram\'irez, I., 2014, ApJL 790, L25
\bibitem[Van Saders \& Pinsonneault (2012)]{vansaders12} Van Saders, J., Pinsonneault, M., 2012, AJ 746, 16

\end{thebibliography}
\end{document}